\documentclass[prl,twocolumn,superscriptaddress,showpacs]{revtex4}
\usepackage{amsmath,amssymb,graphicx}

\def\ketpsi{\vert \psi \rangle}
\def\brapsi{\langle \psi \vert}
\def\ketpsi0{\vert \psi_0 \rangle}
\def\brapsi0{\langle \psi_0 \vert}

\input{tcilatex}

\begin{document}

\title{Asymmetric quantum error correcting codes}
\author{Lev Ioffe}
\affiliation{Center for Materials Theory, Department of Physics and Astronomy, Rutgers University 
136 Frelinghuysen Rd, Piscataway NJ 08854 USA}
\author{Marc M\'ezard}
\affiliation{CNRS; Univ. Paris-Sud, UMR 8626, LPTMS, Orsay Cedex, F-91405 FRANCE}
\date{\today}

\begin{abstract}
The noise in physical qubits is fundamentally asymmetric: in most devices,
phase errors are much more probable than bit flips. We propose a quantum
error correcting code which takes advantage of this asymmetry and shows good
performance at a relatively small cost in redundancy, requiring less than a
doubling of the number of physical qubits for error correction.
\end{abstract}

\maketitle

\emph{Introduction. }The quest of a quantum computer has stimulated a lot of
interesting developments in recent years. However, despite a remarkable
progress, all \ physical devices realized so far do not allow to build even
a very small computer. One crucial aspect is the noise control. Quantum
computing faces two antogonist constraints: one should be able to manipulate
and address the results of a computation, and at the same time one must keep
the noise level low. While some hardware architecture may help to achieve
this compromise, it is clear now that there will never exist a quantum
computer without efficient quantum error correction (QEC).

The basic principles of QEC have been written down in \cite%
{shor,steane,calderbank_shor,steane2}, and a number of QEC codes have been
developed since then \cite{CalRaiShoSlo,kitaev}. However, most of them
require in practice a high level of redundancy (in coding language, a low
rate): the number of physical qubits needed to effectively protect one
logical qubit is large. Generally one expects that a higher rate might be
achievable with good codes when the length of the information block is
large: in this limit, the uncorrected errors correspond to a correlated flip
of a large number of physical bits, the probability of which gets
exponentially small. The classical coding theory shows the existence of
codes that become ideal (saturate the Shannon limit) when the size of the
block tends to infinity. Furthermore, recent progress on so-called low
density parity check (LDPC) codes has revived older ideas by Gallager \cite%
{gallager} that produce efficient algorithms for a fast decoding of codes
with performance close to Shannon's limit \cite%
{MacKaybook,richardsonurbankebook,IEEE_special_issue}. The generalization of
these classical schemes for quantum error correction is made very difficult
by the requirement that a quantum scheme should correct two types of errors:
bit flips as well as phase errors. Another important constraint is the
difficulty to perform operations concurrently on the same bit: an efficient
error correction scheme should involve a relatively small ($o(N)$) read-out
operations on each bit. Here we propose a new family of QEC codes which work
at relatively low redundancy (typically $\leq 2-3$ physical qubits for one
logical qubit), can correct many mistakes and allow parallelization. So far,
the main attempt at finding such codes is the work \cite{McKay}. It uses
so-called self-dual codes which are taylored to deal with a noise which is
symmetric in all channels. We argue that in the physical devices conceived
so far, the noise is typically asymmetric (a phase error is much more
probable than a bit flip), and one can exploit this asymmetry to develop
more efficient QEC codes. The construction that we propose here makes use of
two standard classical codes which are among the most efficient ones: it
handles the relatively rare bit errors through a Bose Chaudhuri Hocquenghem
(BCH) code \cite{bch} and the more frequent phase errors through a LDPC
code, with performances close to those of the most powerful random LDPC codes%
\cite{IEEE_special_issue}.

\emph{Physical noise} The level of the noise in a single physical bit is
conveniently characterized by the relaxation time, $T_{1}$, and dephasing
time $T_{2}$, the two parameters that enter Bloch equation for a single bit
(spin) dynamics. Because the relaxation always implies dephasing, the
dephasing rate $1/T_{2}$ has a contribution from the relaxation processes
and a pure dephasing: $1/T_{2}=1/(2T_{1})+\Gamma _{\phi }$. Generally, there
are many ways to control the relaxation rate: first, the relaxation between
two states with energy difference $\Delta E$ requires a transfer of energy
to the environment, the amplitude of which becomes smaller as $\Delta
E\rightarrow 0$. Furthermore, in many physical implementations these two
states are separated by a large barrier that makes transitions between them
rare. The situation is completely different with the dephasing rate $\Gamma
_{\phi }$ which is physically due to the fluctuations of $\Delta E$ with
time. All low frequency processes contributing to the $\Delta E(t)$
dependence result in the decrease of $\left\langle \exp (-i\int \Delta
E(t)dt\right\rangle $ correlator, i.e. lead to the dephasing. In this
respect, a particularly damaging effect comes from omnipresent $1/f$ noise.
Thus, it is not surprising that in almost all devices studied so far, the
relaxation rate can be made much slower than the dephasing: in a typical NMR
device $T_{1}\sim 10-100s$ while $T_{2}\sim 1s$ \cite{Vandersypen2004}, in
superconducting phase qubits $T_{1}\sim 10\mu s$ while $T_{2}\sim 100ns$ 
\cite{Bertet2005}, in superconducting charge qubit $T_{1}\sim 100ns$ while $%
T_{2}\sim 1ns$ \cite{Astafiev2004} and finally for spin dots $T_{1}\sim 1\mu
s$ while $T_{2}\sim 10ns$ \cite{Elzerman2004,Kato2004}.

In the following we shall therefore assume that in physical qubits the noise
is strongly asymmetric. Specifically, we study a noise channel defined as
follows. Noise acts independently on each bit. It induces a bit flip with
probability $p_x$, and independently it induces a phase flip with
probability $p_z$. The original state of the system, $\vert \psi_0 \rangle$,
is thus changed to $\vert \psi \rangle=\prod_{i}\left[ \left(\sigma_z^i%
\right)^{m_i} \left(\sigma_x^i\right)^{n_i} \right] \vert \psi_0 \rangle $
with probability $p_z^{\sum_i n_i} (1-p_z)^{N-\sum_i n_i} p_x^{\sum_i m_i}
(1-p_x)^{N-\sum_i m_i}$, where $m_i,n_i \in \{0,1\}$. The channel acts on
bit $i$ by applying an operator $U_i\in\{\mathcal{I},\sigma_x^i,\sigma_z^i,
\sigma_x^i\sigma_z^i\}$.

\emph{CSS codes.} Our family of codes is of the CSS type \cite%
{calderbank_shor,steane2}. It consists of two independent encoding/decoding
devices dealing separately with bit and phase flips, for a string of $N$
physical qubits. It uses $M_z$ '$z$-checks' and $M_x$ '$x$-checks'. The $a$%
-th $z$-check is defined by a set $V(a)\in \{1,\dots,N\}$ and by the
operator $C_a^z = \prod_{i \in V(a)} \sigma_i^z$. Similarly, the $a$-th $x$%
-check is defined by a set $W(a)\in \{1,\dots,N\}$ and by the operator $%
C_a^x = \prod_{i \in W(a)} \sigma_i^x$.

By construction, the $z$-checks and $x$-checks all commute with each other,
and the original state $|\psi _{0}\rangle $ is an eigenstate of all the
operators $C_{a}^{z},C_{a^{\prime }}^{x}$ with eigenvalue $1$. As $U_{i}$
either commutes or anticommutes with these check operators, the
noise-perturbed state $|\psi \rangle $ is an eigenstate of the operators $%
C_{a}^{z},C_{a^{\prime }}^{x}$. The decoding operation consists of three
steps: (i) measure the eigenvalues of the check operators, ii) infer from
these eigenvalues what was the corrupting operator, (iii) apply the
correction operator. In more detail:

Step (i): The $a$-th $z$-syndrom is defined as the number $u_{a}\in \{0,1\}$
such that $C_{a}^{z}|\psi \rangle =(1-2u_{a})|\psi \rangle $. Similarly, the 
$a$-th $x$-syndrom is defined as the number $v_{a}\in \{0,1\}$ such that $%
C_{a}^{x}|\psi \rangle =(1-2v_{a})|\psi \rangle $.

Step (ii): From the $z$-syndroms $\{u_{a}\}$, $a\in \{1,\dots ,M_{z}\}$, we
compute $N$ numbers $\{m_{1}^{\prime },\dots ,m_{N}^{\prime }\}$ such that,
for each $a\in \{1,\dots ,M_{z}\}$: $\sum_{i\in V(a)}m_{i}^{\prime
}=u_{a}(mod\ 2)$, with the smallest possible number of $m^{\prime }$s equal
to $1$. From the $x$-syndroms $\{v_{a}\}$, $a\in \{1,\dots ,M_{x}\}$, we
compute $N$ numbers $\{n_{1}^{\prime },\dots ,n_{N}^{\prime }\}$ such that,
for each $a\in \{1,\dots ,M_{x}\}$: $\sum_{i\in W(a)}n_{i}^{\prime }=v_{a}(mod\ 2)$,
with the smallest possible number of $n^{\prime }$s equal to $1$.

Step (iii): generate $|\psi ^{\prime }\rangle =\prod_{i=1}^{N}\left[ (\sigma
_{x}^{i})^{n_{i}^{\prime }}(\sigma _{z}^{i})^{m_{i}^{\prime }}\right] |\psi
\rangle $. If the error correction is successfull, one should find $|\psi
^{\prime }\rangle =|\psi _{0}\rangle $.

A CSS code is thus characterized by the sets $V(a)$ and $W(a)$ defining the
checks. In building such a code, one must ensure that all check operators
commute. This imposes that $\forall a\in \{1,\dots ,M_{z}\}$, $\forall
a^{\prime }\in \{1,\dots ,M_{x}\}$, the cardinal of $|V(a)\cup W(a^{\prime
})|$ be even. It is useful to define the parity check matrices of the two
codes. The matrix $H^{z}$ is a $M_{z}\times N$ matrix with entries in $%
\{0,1\}$, defined by $H_{ai}^{z}=1$ if and only if $i\in V(a)$. Similarly, $%
H^{x}$ is the $M_{x}\times N$ matrix defined by $H_{ai}^{x}=1$ if and only
if $i\in W(a)$. The commutativity condition is satisfied when $H^{z}\left(
H^{x}\right) ^{T}=0$ (using Boolean algebra, i.e. mod(2) additions). The $z$%
-codewords are strings of $N$ bits $x_{i}\in \{0,1\}$ such that, $\forall a$%
, $\sum_{i}H_{ai}^{z}x_{i}=0(mod\ 2)$. Any $x$-check $a$ defines a $z$%
-codeword through $x_{i}=1$ if $i\in W(a)$, and $x_{i}=0$ otherwise.
Similarly, $z$-checks define $x$-codewords. Most of the research on QEC so
far has focused on the design of relatively small codes with good distance
properties. If for instance all pairs of $x$-codewords are at a Hamming
distance $\geq 2d+1$, the code will correct any set of $\leq t$ flip errors.
While this suggest to build codes which maximize the smallest distance
between codewords, this strategy is not necessarily optimal when dealing
with large blocklength ($N\gg 1$). Instead, what is practically required is
that the probability of an error is small and it turns out that the best
classical codes often have (very rare) pairs of codewords which are pretty
close to each other\cite{IEEE_special_issue}. We shall use this approach to
construct our $x$-checks.

\emph{$z$-checks: BCH code} Our $z$-code is an efficient classical
construction, a binary primitive BCH code (see ref. \cite{bch_rev} for an
extended presentation). The code depends on two parameter $m,t$. The first
one determines the Galois field $GF(2^{m})$ which is used, and the number $t$
is equal to the number of errors (phase flips) that the code can correct.
The number of variables (and therefore the number of qubits) is given by $%
N=2^{m}-1$. If $\alpha $ is a primitive element of the field $GF(2^{m})$,
the powers $\alpha ^{r},\ r\in \{1,\dots ,N\}$ are $N$ distinct elements of
the field, building a cyclic group under multiplication. At the same time, $%
GF(2^{m})$ is a vector space of dimension $m$ over $GF(2)$: every element $%
\alpha ^{r}$ can be decomposed as $\alpha ^{r}=\sum_{p=0}^{m-1}\gamma
_{rp}\alpha ^{p}$, where the coefficients $\gamma $ are in $\{0,1\}$. The
check matrix $H$ of the code is defined as 
\begin{equation}
H=\left[ 
\begin{array}{ccccc}
1 & \alpha & \alpha ^{2} & \dots & \alpha ^{N-1} \\ 
1 & (\alpha ^{3}) & (\alpha ^{3})^{2} & \dots & (\alpha ^{3})^{N-1} \\ 
1 & (\alpha ^{5}) & (\alpha ^{5})^{2} & \dots & (\alpha ^{5})^{N-1} \\ 
\dots & \dots & \dots & \dots & \dots \\ 
1 & (\alpha ^{2t-1}) & (\alpha ^{2t-1})^{2} & \dots & (\alpha ^{2t-1})^{N-1}%
\end{array}%
\right]
\end{equation}%
This matrix can be seen either as a $t\times N$ matrix with elements in $%
GF(2^{m})$, but another interpretation is also useful. If we write each
element $\alpha ^{r}$ of $H$ as the $m$ component vector $\left( 
\begin{array}{c}
\gamma _{r0} \\ 
\dots \\ 
\gamma _{r(m-1)}%
\end{array}%
\right) $, we obtain the $t m\times N$ parity check matrix $H^z$ with
entries in $GF(2)=\{0,1\}$. Therefore $M_z=t m$. BCH decoding relies on
algebraic properties which are most easily written in terms of polynomials.
Here we shall just present the basic result in the case $t=2$. If two of the 
$N$ bits are flipped by noise, and these indices correspond to the elements
of $GF(2^{m})$ called $\beta _{1},\beta _{2}$, the check matrix $H$ ,
applied to the error vector, gives two syndromes $\zeta _{1}=\beta
_{1}+\beta _{2}$ and $\zeta _{3}=\beta _{1}^{3}+\beta _{2}^{3}$. Decoding
consists in finding $\beta _{1},\beta _{2}$ given $\zeta _{1},\zeta _{2}$.
It is easily seen that this system has a unique solution in $GF(2^{m}) $ (up
to the permutation of $\beta _{1}$ and $\beta _{2}$): the code with $t=2$
corrects exactly any set of $\leq 2$ errors. The same construction works for
arbitrary $t $, and good decoding algorithms exist: the code corrects any
set of $\le t$ errors. In practice we have used the Berlekamp algorithm \cite%
{bch_rev}, adapting some software available from \cite{Morelos}.

\emph{Generation of the $x$ checks: LDPC code}. Some BCH codes are
self-dual; in such a case one gets a quantum code using $H^x=H^z$ \cite%
{steane_bch}. But in order to get a much better performance (for large $N$)
on the $x$-channel, we prefer to use a code as close as possible to the
random LDPC codes. The commutation of the $x$ and $z$ checks is obtained by
the following procedure. Given a BCH code with parameters $m,t$, we can
generate a $x$-check $a$ with any degree $n\geq 2t+1$ using a variant of its
standard decoding algorithm. The first $n-t$ elements of $W(a)$ are chosen
as a random subset of $\{1,\dots ,N\}$ with distinct elements, taken
uniformly among all such subsets. Let us call $\beta _{1},\dots ,\beta_{n-{t}%
}$ the corresponding elements of $GF(2^{m})$. We look for the remaining $t$
elements which are solutions of the decoding equations 
\begin{eqnarray}
\forall s \in\{1,\dots,t\}:\ \ \sum_{r=1}^t
\left(\beta_{n-t+r}\right)^{2s-1}=- \sum_{r=1}^{n-t}
\left(\beta_{r}\right)^{2s-1}\ .
\end{eqnarray}%
Provided that the solution of these equations exists (which happens with
probability $1/t!$) the elements $\beta _{d-t+1}\ldots \beta _{d}$ can be
found using any standard BCH decoding algorithm, like Berlekamp's one. The
indices corresponding to the elements $\beta _{1},\dots ,\beta _{n-t}\ldots
\beta _{n}$ form the subset $W(a)$ defining the $a$-th $x$-check. As $\beta
_{1},\dots \beta _n$ is a codeword of the BCH code, the commutativity
condition is satisfied.

Clearly, the indices in $V(a)$ do not form a random subset of size $n$.
However, if the map used in generating $\beta _{n-t+1}\ldots \beta _{n}$
from $\beta _{1},\dots ,\beta _{n-t}$ is chaotic enough (we shall refer to
this hypothesis as the 'chaos hypothesis' in the following), one can hope to
generate a set of $x$-checks with performances close to the ones of
classical random LDPC codes. This is what we have found numerically. In
practice, for a given value of $t$, we generate a large enough pool of
parity checks, all having degree $n=2t+1$. From this pool, we select a
number $M_{z}$ of checks, in such a way that the degrees of the variables in
the corresponding factor graph has a narrow distribution. This is done by
the following inductive procedure. At each step we order the remaining
(unused) set of checks by their 'quality' which is defined as the number of
minimal degree variables that would be affected by addition of this check.
We then add one (randomly chosen) check of the highest 'quality' and repeat
the procedure.

The practical decoding of our LDPC code uses the standard 'belief
propagation' (BP) algorithm\cite{MacKaybook,richardsonurbankebook}, a
message passing algorithm which is equivalent to an iterative solution of
Bethe equations.

\emph{Performance.} An important parameter of the code is its degree of
redundancy. We have checked that the various checks are generically linearly
independent, so the $z$-rate (resp. $x$ rate) is obtained as $R_z=1-\frac{M_z%
}{N}$ (resp $R_x=1-\frac{M_x}{N}$) and the quantum rate of the code is $R= 1
-\frac{M_x+M_z}{N}$.

The error correction ability depends on the channel. In the $z$-channel (bit
flip errors), by construction, the BCH code is able to decode up to $t$
errors. Therefore the probability of error in decoding this channel is 
\begin{equation}
P_{\text{err}}^z= \sum_{j=t+1}^N \binom{N}{j} p_z^j (1-p_z)^{N-j}\ ,
\label{Perrz}
\end{equation}
which is well approximated, for the small values of $p_z$ which interest us
here, by $1-e^{-Np_z} \sum_{j=0}^t (Np_z)^j/j!$.

Let us now turn to the $x$-channel. The performance of BP decoding for
random LDPC codes can be studied analytically in the limit of large
blocklength \cite{IEEE_special_issue}. Within the chaos hypothesis, one
could thus derive the threshold for zero error decoding in the large $N$
limit. However in practice we are interested in not-too-large values of $N$.
We have thus tested numerically the BP decoding of our $x$-code.

The simulation is run as follows. We fix an 'acceptable' value of the block
error $P_{\text{block}}$ for decoding $N$ bits, both in the $x$ and in the $%
z $-channel, in practice $P_{\text{block}}=10^{-4}$. For given values of $N$
(or $m$) and $t$, eq.(\ref{Perrz}) gives the noise level $p_{z}$ that can be
corrected in the $z$-channel, and the channel asymmetry gives the ratio $%
p_{z}/p_{x}$. We then test various $x$-codes, varying $M_{x}$ until the
block error in the $x$-channel is less than $P_{\text{block}}$. Results are
summarized in the following table, which studies asymmetries $%
p_{z}/p_{x}=0.01,\,0.1$.

\begin{tabular}{||c|c||c|c|c||c|c|c|c||}
\hline
$m$ & $t$ & $p_z$ & $M_z$ & $Q_z$ & $p_x$ & $M_x$ & $Q_x$ & $Q$ \\ \hline
10 & 2 & $8.40 \; 10^{-5}$ & 20 & .980 & $8.4\; 10^{-3}$ & 563 & .45 & .43
\\ 
10 & 3 & $2.26 \; 10^{-4}$ & 30 & .971 & $2.26 \; 10^{-2}$ & 460 & .55 & .52
\\ 
10 & 4 & $4.34 \; 10^{-4}$ & 40 & .961 & $4.34 \; 10^{-2}$ & 530 & .48 & .44
\\ 
10 & 3 & $2.26 \; 10^{-4}$ & 30 & .971 & $2.26 \; 10^{-3}$ & 460 & .55 & .52
\\ 
10 & 4 & $4.34 \; 10^{-4}$ & 40 & .961 & $4.34 \; 10^{-3}$ & 344 & .66 & .62
\\ 
10 & 5 & $6.98 \; 10^{-4}$ & 50 & .951 & $6.98 \; 10^{-3}$ & 271 & .73 & .69
\\ 
10 & 6 & $1.01 \; 10^{-3}$ & 60 & .941 & $1.01 \; 10^{-2}$ & 285 & .72 & .66
\\ 
12 & 3 & $5.66 \; 10^{-5}$ & 36 & .991 & $5.66 \; 10^{-3}$ & 1577 & .61 & .61
\\ 
12 & 4 & $1.08 \; 10^{-4}$ & 48 & .988 & $1.08 \; 10^{-2}$ & 1378 & .66 & .65
\\ 
12 & 5 & $1.74 \; 10^{-4}$ & 60 & .985 & $1.74 \; 10^{-2}$ & 1189 & .71 & .69
\\ 
12 & 6 & $2.52 \; 10^{-4}$ & 72 & .982 & $2.52\; 10^{-2}$ & 1191 & .71 & .69
\\ \hline
\end{tabular}

\begin{figure}[tbp]
\begin{center}
\includegraphics[angle=-90,width=.99\linewidth]{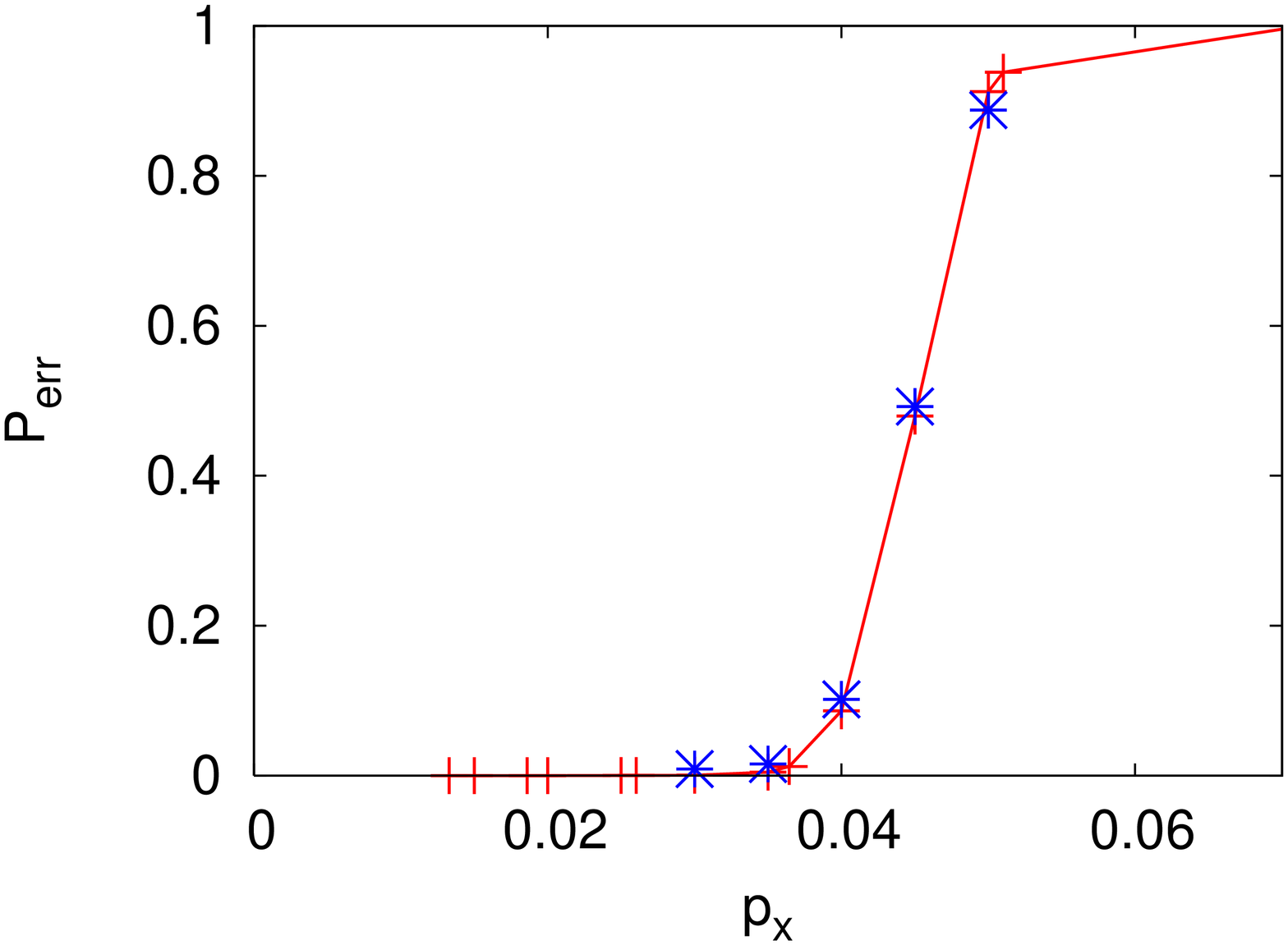} \put(-185,-17){%
\includegraphics[width=0.34\linewidth,angle=-90]{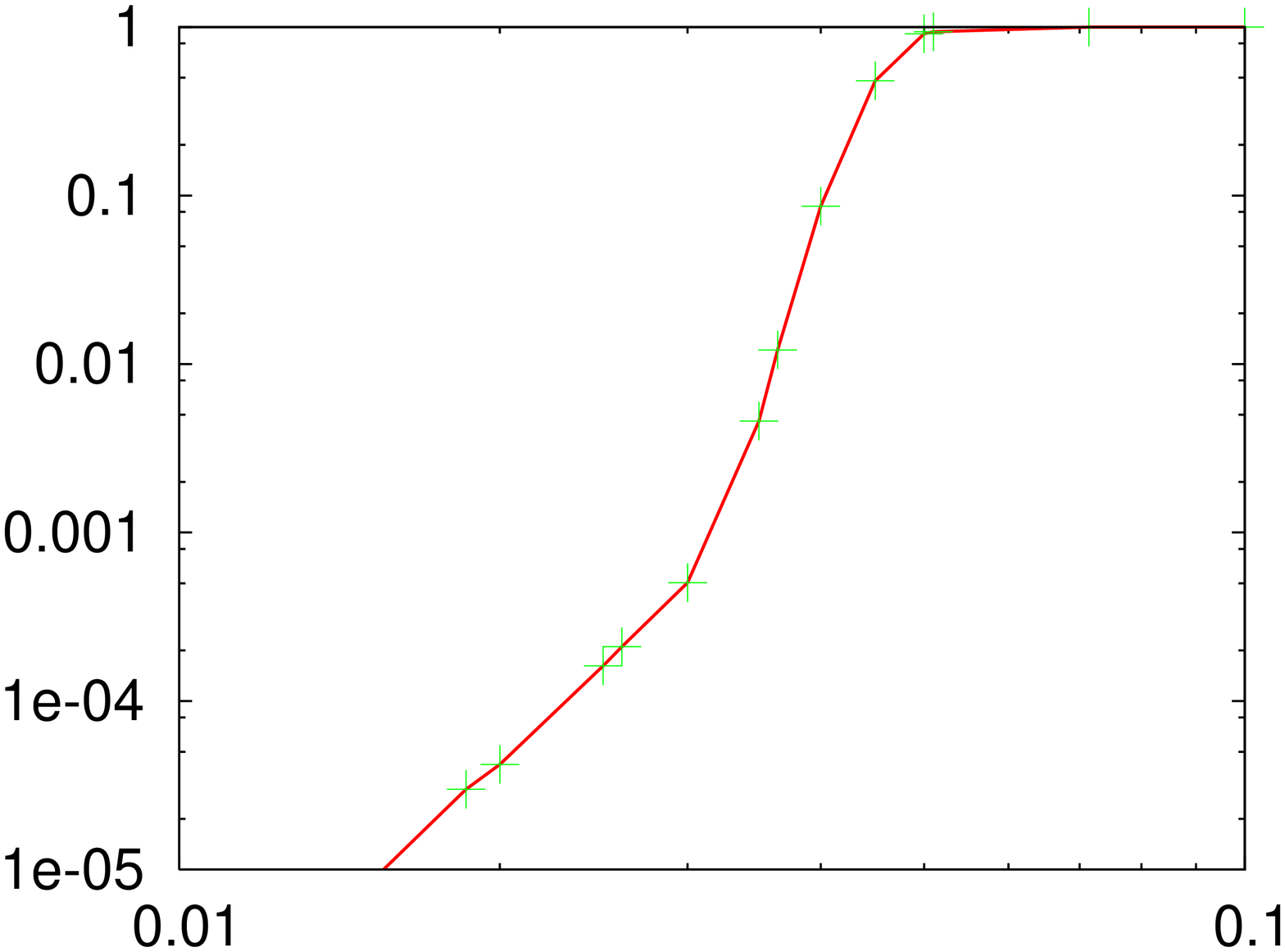}}
\end{center}
\caption{ $+$: block error in the $x$-channel, $P_{\text{err}}^{x}$, versus
the phase error probability $p_{x}$, for the code with $m=12$, $t=4$, $%
M_{x}=1378$. The line is a guide to the eye. Also shown is the same curve
for a random LDPC code ($\times $). The inset gives the same data in a
log-log plot}
\label{fig:perr_px}
\end{figure}

We see that large enough codes provide a good performance. For instance, $%
m=12,t=6$, code with $N=4095$ qubits is able to correct a noise level of $%
p_{z}=2.5\;10^{-4}$ in the $z$ channel and $p_{x}=2.5\;10^{-2}$ in the $x$
channel with block error probability smaller than $10^{-4}$. Notice that for these
values of $p_{z}$, $p_{x}$, the probability of a block error \emph{without }%
any error correction would be $1-(1-p_{z,x})^{N}$, giving $.63$ for the $z$%
-channel and $1$ for the $x$-channel. Figure (\ref{fig:perr_px}) gives the
block error in the $x$-channel, $P_{\text{err}}^{x}$, versus the phase error
probability $p_{x}$, for one given code. 

\emph{Conclusions.} We have provided an explicit construction of quantum
codes with rates $Q\sim 0.5$ that are able to correct a few errors in one
channel (bit flips) and have close to optimal peformance in another (phase
errors), together with efficient decoding procedures.
 One important aspect of these codes is the fact that the number of operations to be done on one
given bit is much smaller than $N$. In the $z$-channel this is due to the
fact that we use a small value of $t$, in the $x$-channel it is due to the
intrinsic low density of the code. We believe that these codes might be
quite useful for the realistic physical implementation of quantum memory. We
have not investigated the possibility of using them for fault tolerant
quantum computation, this is the subject of the future research.

\emph{Acknowledgments.} We thank J.S. Yedidia for interesting discussions.
This work has been supported in part by the EC grants 'Stipco',
HPRN-CT-2002-00319, 'Evergrow', IP 1935 in the FET-IST programme and NSF DMR
0210575. LI thanks LPTMS for the hospitality that made this work possible.


\end{document}